\documentclass[reqno,12pt]{amsart}

\usepackage{amsmath}
\usepackage{amssymb}
\usepackage[pdftex]{graphicx}
\usepackage{xy}
\usepackage{pdfpages}
\usepackage{fancyhdr}

\usepackage{hyperref}

\theoremstyle{plain}

\newcommand{\HRule}{\rule{\linewidth}{0.5mm}}

\author{Michael Robinson}
\address{Department of Mathematics and Statistics\\American University\\4400 Massachusetts Ave NW\\Washington, DC 20016}
\email{michaelr@american.edu}



\title{Doppler Tracking of the Artemis II Mission (and Other Spacecraft)}

\begin{document}

\begin{titlepage}

\begin{center}

\thispagestyle{fancy}
\textsc{\large Technical Report}\\[0.2cm]

\HRule \\[0.2cm]
{ \huge \bfseries Doppler Tracking of the Artemis II Mission (and Other Spacecraft)}\\[0.4cm]

\HRule \\[0.4cm]

\begin{minipage}{0.4\textwidth}
\begin{flushleft}
\emph{Technical contact:}\\
Prof.~Michael \textsc{Robinson}\\
Mathematics and Statistics\\
American University\\
4400 Massachusetts Ave NW\\
Washington, DC 20016\\
(202)885-3681\\
michaelr@american.edu\\[0.2cm]
\end{flushleft}
\end{minipage}
\begin{minipage}{0.4\textwidth}
\begin{flushleft}
      \emph{NASA POC:}\\
Dr.~Marta \textsc{Sheldon}\\
Associate Branch Head Code 524\\
SCaN Commercialization \\
Innovation and Synergies (CIS)\\
Goddard Space Flight Center\\
O: 301-614-5624\\
M: 757-894-1005\\
marta.b.shelton@nasa.gov\\[0.2cm]
\end{flushleft}
\end{minipage}

\vspace{2cm}

\large \centerline {Ankur Purao, Ella Bianco, Elizabeth Baganz, Nicholas Hagg, Michael Robinson}

\vspace{2cm}

\includegraphics[height=0.5in]{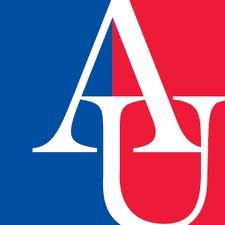}\\
\Large American University

\vspace{2cm}

Spring 2026\\[1cm]

\vfill

\end{center}

\end{titlepage}

\newpage

\tableofcontents

\newpage

\chead{}
\rhead{}
\lfoot{}
\cfoot{}
\rfoot{\thepage}

\setcounter{page}{1}

\section{Technical objective}

\begin{figure}[!htbp]
  \begin{center}
    \includegraphics[width=4in]{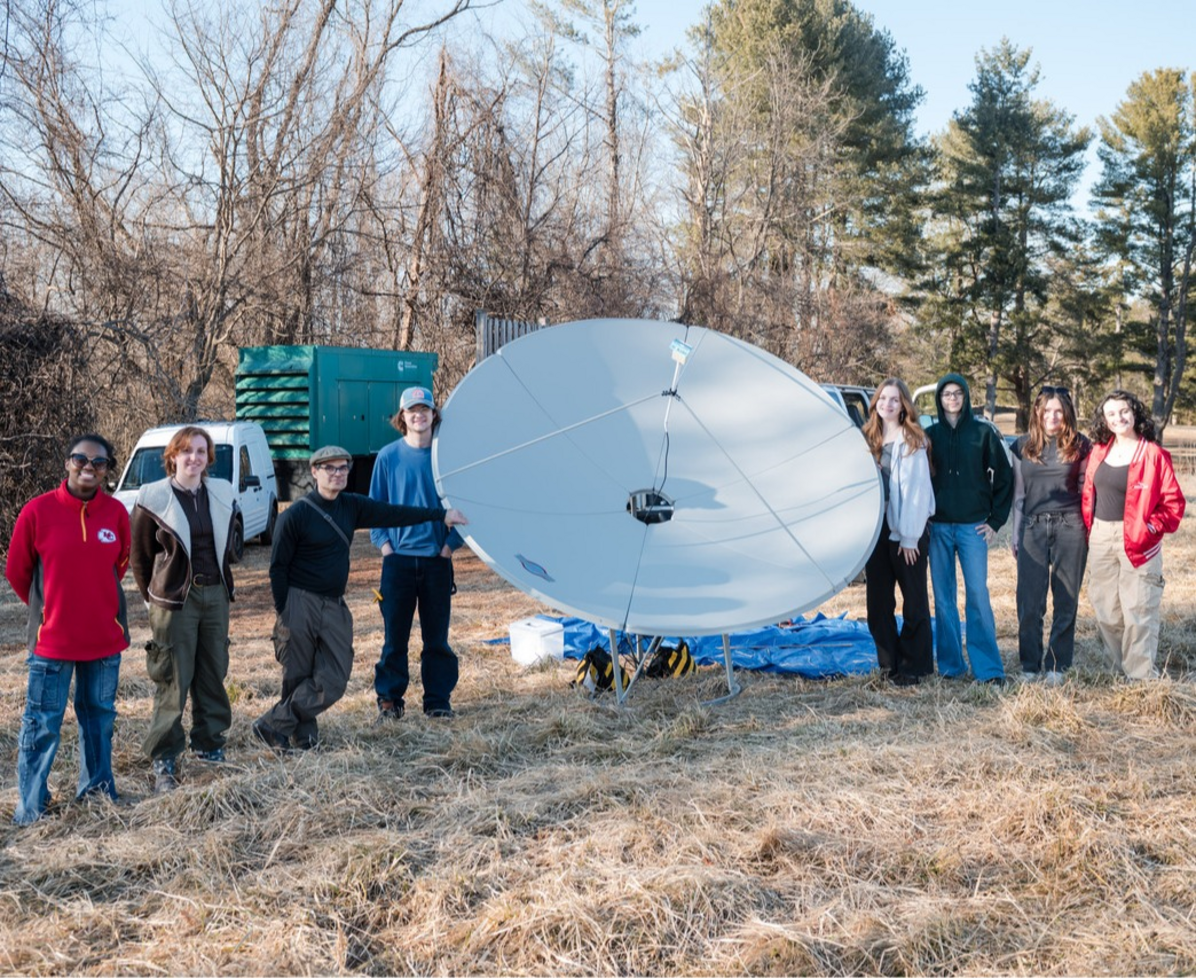}
    \caption{The American University Artemis II Tracking Team (Photo by Nikolai Roster, CAS.)}
    \label{fig:group_photo}
  \end{center}
\end{figure}

This report describes an American University (AU) student-faculty project to track the Orion spacecraft ``Integrity'' during the Artemis II mission,
along its Earth-to-Moon trajectory.
The Orion spacecraft has an S-band communications transmitter used for space-to-Earth messages.
The S-band trasmitter uses an orthogonal quadrature phase shift keying (OQPSK) digital mode,
from which Doppler frequency measurements can be derived.

Although our main goal is to collect accurate Doppler tracking data,
there is also substantial educational value that we hope to glean from this project.
The project is primarily staffed and supported by the AU Physics Department, which is an entirely undergraduate department.
The Physics Department specializes in introducing undergraduate physics majors to physics as a whole and providing them with hands-on experience.
At the same time, the department is looking to expand its research opportunities in astrophysics in particular,
and is hoping to use this as a way to introduce undergraduates into experimental astrophysics.
Therefore, while we are hoping to collect relevant data, there is an educational aspect that is extremely important.
Undergraduates did the vast majority of the data collection and will use this experience to guide their career trajectories.
This was, for many of the students on the project, the first time that they conducted actual experimental physics.
For these students, projects such as this are hallmark occasions on their academic journeys for extremely relevant and important research experience.

\subsection{Team previous experience}

\begin{figure}[!htbp]
  \begin{center}
    \includegraphics[width=4in]{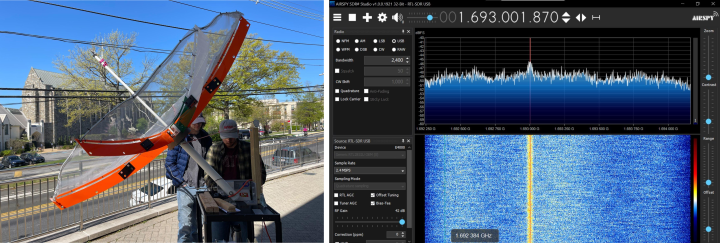}
    \caption{Our team's previous geostationary satellite collection experiment.}
    \label{fig:previous_dish}
  \end{center}
\end{figure}

The group of students on this project already had a significant amount in radio astronomy and S-Band detection is particular.
Figure \ref{fig:previous_dish} shows an example student project that was completed in the Spring of 2025,
in which a student built and designed a ground station from scratch in order to receive transmissions from the GOES-19/GOES-EAST geostationary satellite.
The $1.825$ m parabolic dish was designed and made from 3-D printed parts and laser-cut acrylic with mesh zip tied to the frame.
The ground station was located at AU in Washington DC,
so the noise temperature of the location was extremely high due to being in a densely populated urban setting.
Despite that, the system was able to sync with the GOES-19 satellite and receive telemetry data at $1.694$ GHz.

\begin{figure}[!htbp]
  \begin{center}
    \includegraphics[width=4in]{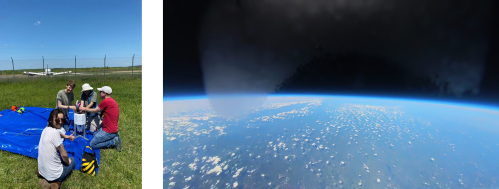}
    \caption{Our team's previous high altitude balloon launch (left) and view from the balloon near zenith at about 100,000 ft (right).}
    \label{fig:balloon}
  \end{center}
\end{figure}

Even before the radio astronomy group was formed,
many of the students and faculty currently involved in it were involved in a 2023--2024 high altitude particle
detection project that successfully launched, tracked, and retrieved a payload that reached an
elevation of $100 000$ ft. The payload was launched in order to collect data on high altitude muon
formation and was launched on April 28th, 2024 from Brookneal-Campbell County Airport,
Virginia. Photos of the launch and pictures taken from the payload are shown in Figure \ref{fig:balloon}.

\begin{figure}[!htbp]
  \begin{center}
    \includegraphics[width=4in]{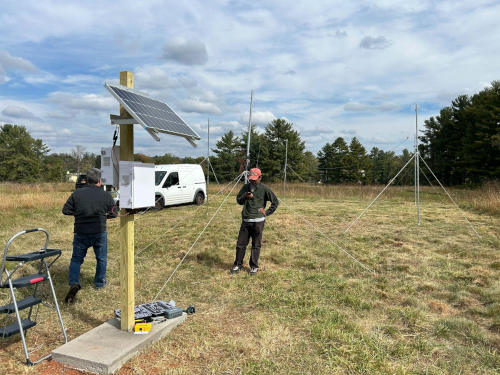}
    \caption{Radio JOVE telescope at the Airlie site.}
    \label{fig:jove}
  \end{center}
\end{figure}

The AU radio astronomy group also recently fielded a radio telescope using a dual dipole antenna (Figure \ref{fig:jove}) in order to measure solar flares.
This telescope uses the Radio JOVE system \url{https://radiojove.gsfc.nasa.gov}.
It is located at the same site at Airlie Farm that was used for the Orion spacecraft tracking experiment (Section \ref{sec:site}).

\section{System design}

\begin{figure}[!htbp]
  \begin{center}
    \includegraphics[width=6in]{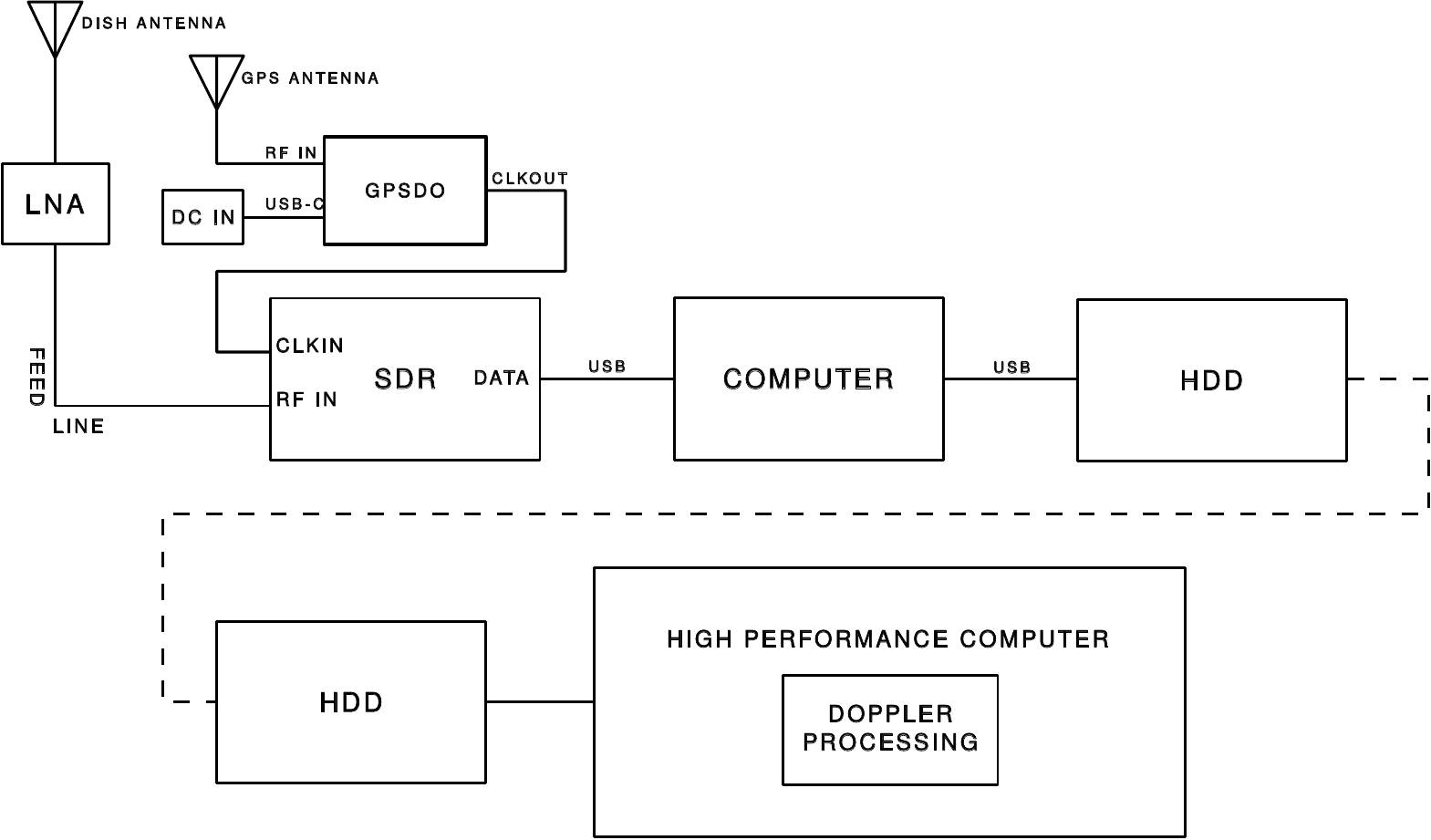}
    \caption{Spacecraft Doppler measuring system block diagram.}
    \label{fig:feedline_schematic}
  \end{center}
\end{figure}

Figure \ref{fig:feedline_schematic} shows the overall system block diagram for the Orion S-band transmitter Doppler measurement project.
Each block is explained in the sections below.

\subsection{Site selection}
\label{sec:site}

\begin{figure}[!htbp]
  \begin{center}
    \includegraphics[width=6in]{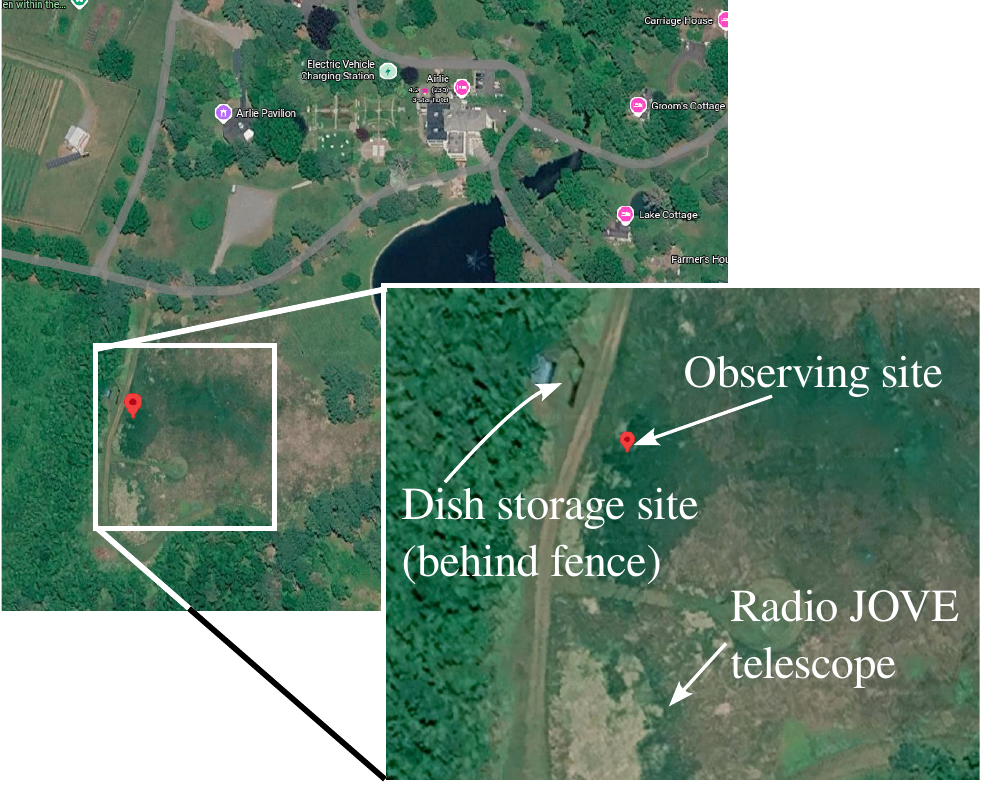}
    \caption{Observing site in context.  Also shown are the locations where the dish is stored when not in use, and the location of the Radio JOVE telescope site.  (Image source: Google Maps)}
    \label{fig:site}
  \end{center}
\end{figure}

Our observing site is located at Airlie Farm in Warrenton, VA.
We chose a site in an open field located at $38.754883^\circ$ N, $77.795939^\circ$ W (see Figure \ref{fig:site}).
Trees obscure elevations below $10^\circ$ but a clear view of much of the sky is available for higher elevations.
This location is adjacent to our existing paired dipole radio telescope.
With a highly directional antenna, terrestial radio interference sources are largely minimized,
but we wanted to ensure that the noise floor for our observations was as close to the thermal background as possible.
Ultimately, as Sections \ref{sec:link_budget} and \ref{sec:data_summary} show,
although the theoretical noise floor is about $-107$ dB, the observed noise floor was $-90$ dB.
The noise figure of the radio at our operating frequency is about $10$ dB \footnote{\url{https://greatscottgadgets.com/2025/12-03-hackrf-pro-receive-sensitivity-and-noise-figure/}}, with approximately $3$ dB\footnote{A guess based upon similar hardware.} contributed by the dish feed assembly low-noise amplifier (LNA),
so the local noise contribution from the site was about $7$ dB, which is quite good.

\subsection{Antenna assembly}

\subsubsection{Parabolic dish}

\begin{figure}[!htbp]
  \begin{center}
    \includegraphics[width=4in]{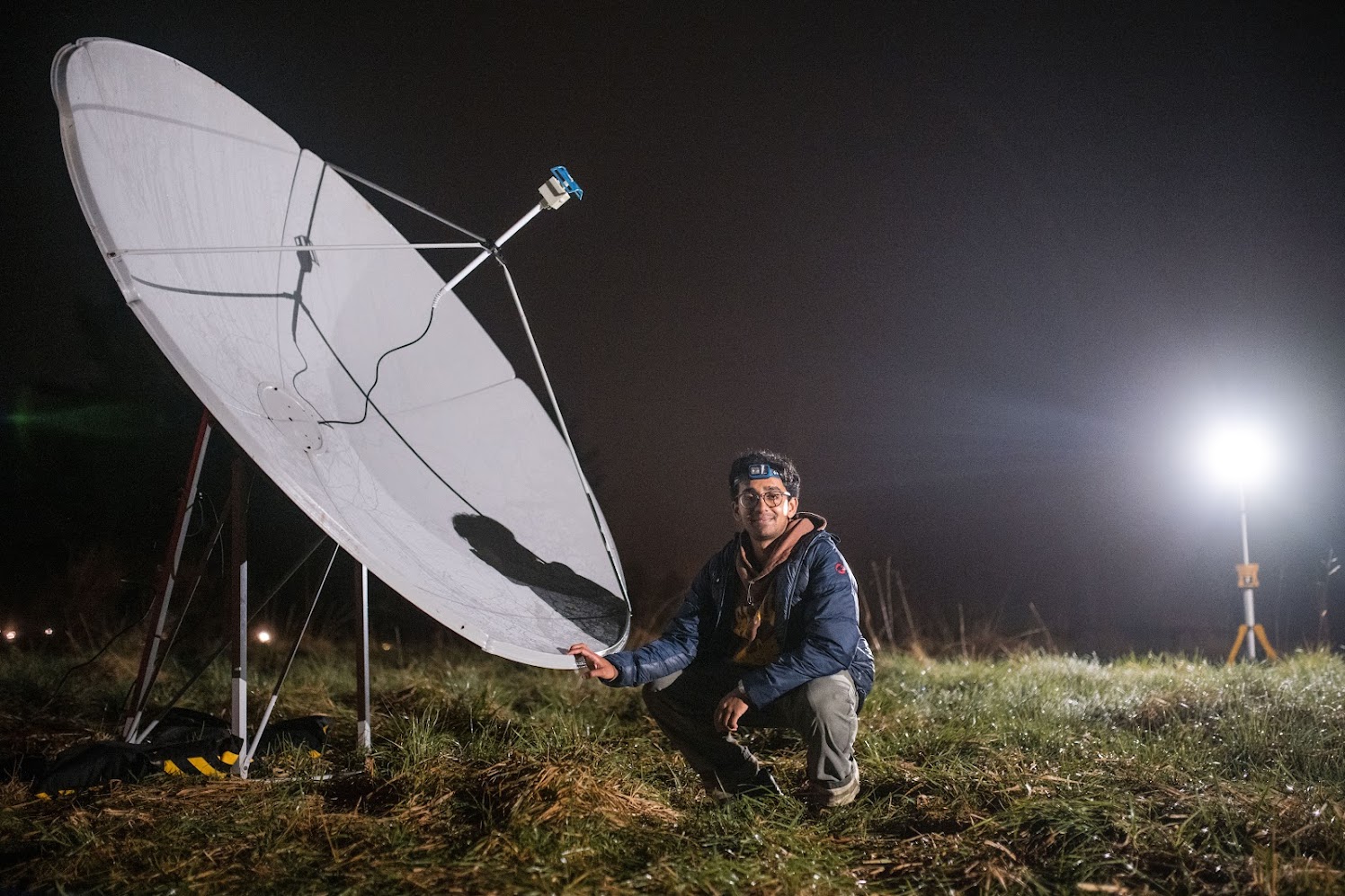}
    \caption{View of the parabolic dish.  (Photo by Nikolai Roster, CAS.)}
    \label{fig:dish}
  \end{center}
\end{figure}

As shown in Figure \ref{fig:dish}, the parabolic dish we used is $2.4$ m in diameter.
This dish was purchased from \url{https://satellitedish.com/},
  and was shipped unassembled on a pallet to Airlie Farm's machine shop.
A team of students assembled the dish and its mounting assembly.

Antenna gain at the operating frequency of $2.2165$ GHz (assuming efficiency of $0.5$, which is perhaps optimistic) is
\begin{equation*}
  10\times \log_{10}(0.5 (\pi \times 2.4 \times 2.2165\times 10^{9}/3\times 10^8)^2) = 32 \text{ dB}.
\end{equation*}

\subsubsection{Mounting and pointing}

\begin{figure}[!htbp]
  \begin{center}
    \includegraphics[width=4in]{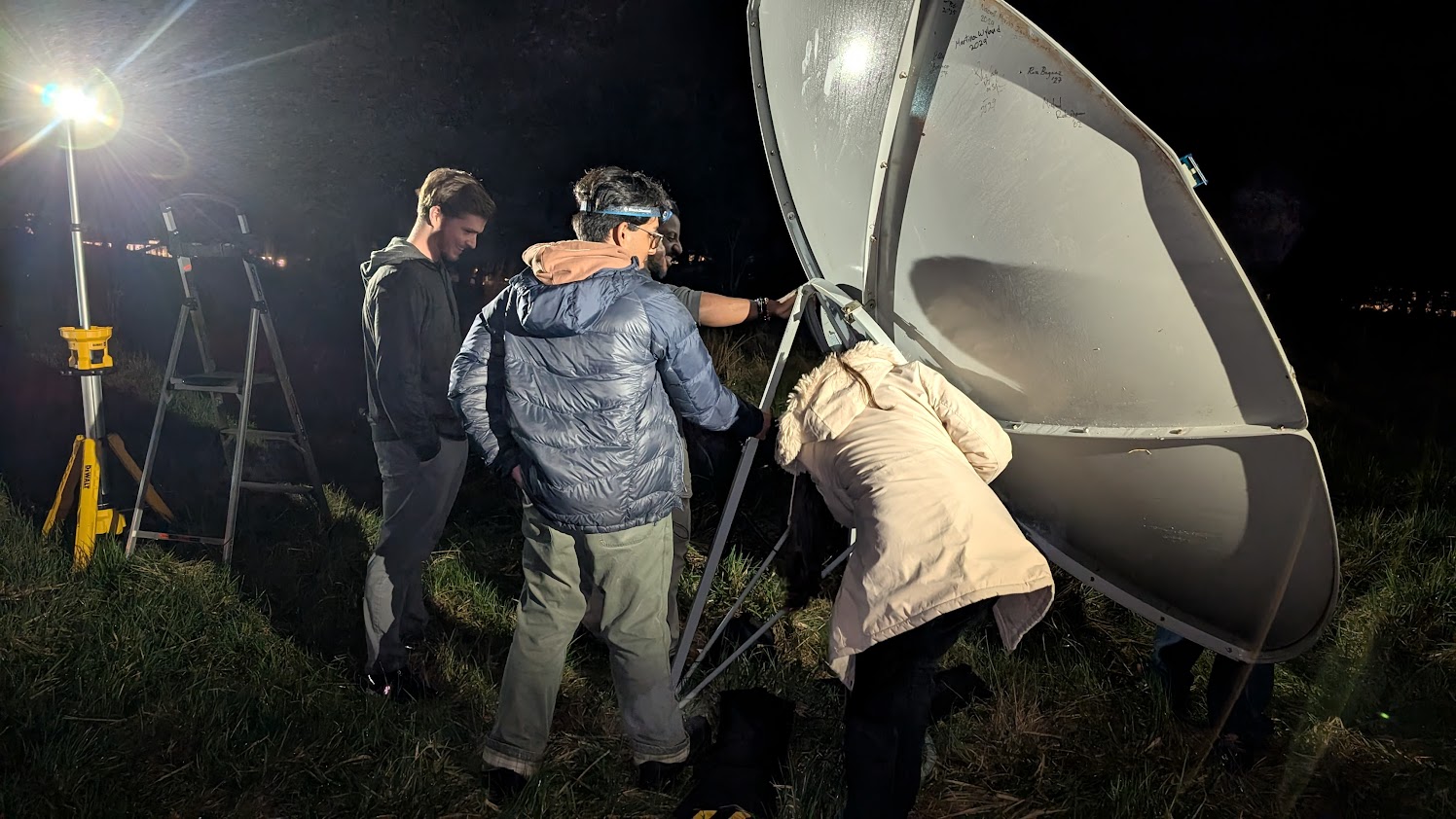}
    \caption{Manually pointing the dish using a commpass and inclinometer.  (Photo by Nikolai Roster, CAS.)}
    \label{fig:pointing}
  \end{center}
\end{figure}

The dish included a rigid metal frame with a locking knob to control elevation.
Since the azimuth and elevation of the Orion spacecraft do not change rapidly over our collection window (Table \ref{tab:pointing}),
we manually adjusted azimuth and elevation at regular intervals throughout the collection.
Figure \ref{fig:pointing} shows an adjustment to the pointing in progress.
One student is responsible for measuring the elevation with a smartphone inclinometer.
Several other students used magnetic compasses to establish the azimuth.
Since the dish is ferromagnetic, several compass readings from several locations were necessary to ensure reliable azimuth measurements.
Once the correct pointing was established, the dish was anchored by placing several sandbags on the base of the mounting frame.

\subsubsection{Feedpoint assembly}

\begin{figure}[!htbp]
  \begin{center}
    \includegraphics[width=4in]{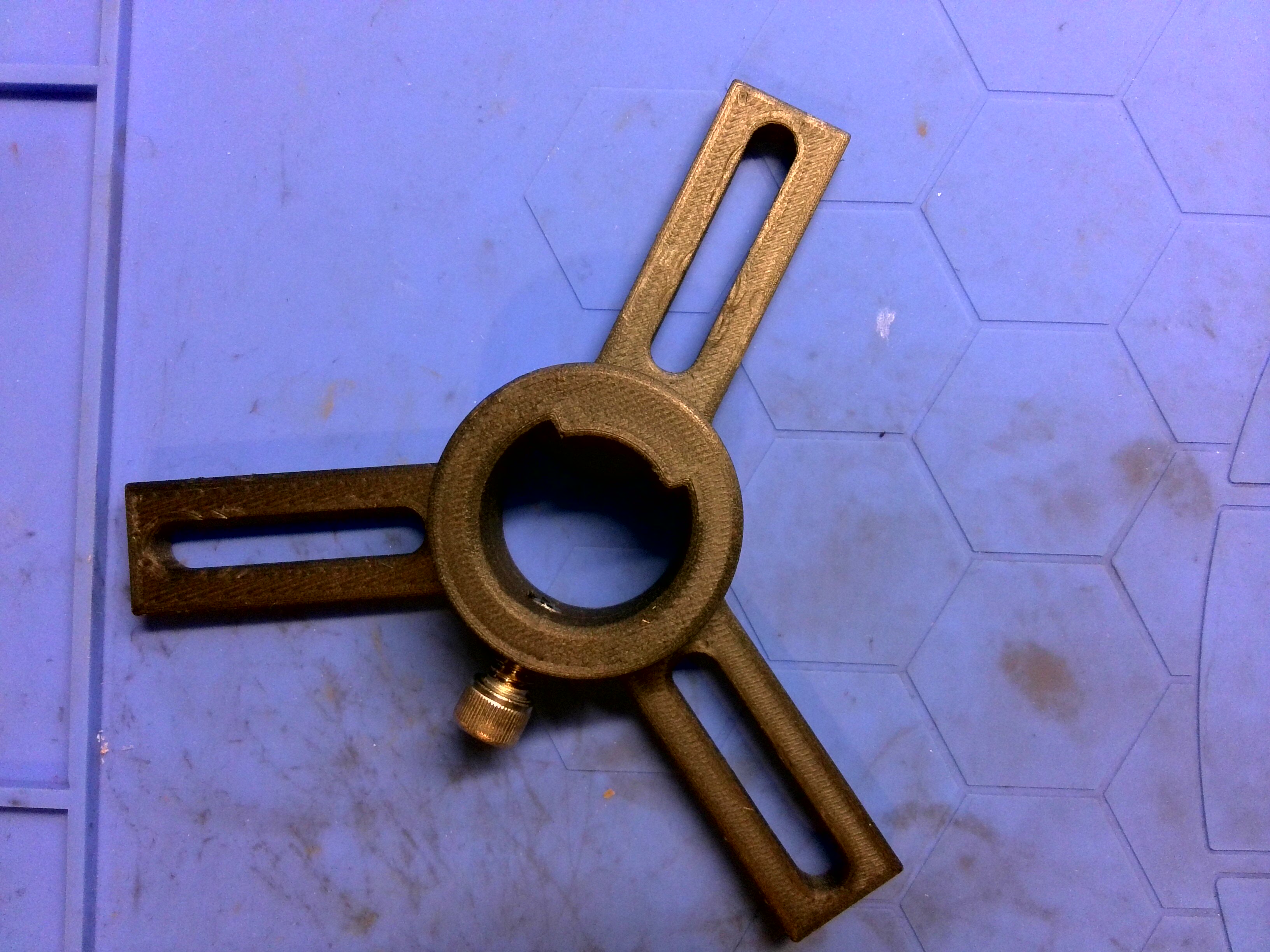}
    \caption{Feedpoint mounting bracket (top view)}
    \label{fig:bracket}
  \end{center}
\end{figure}

The parabolic dish has three mounting arms that are used to hold a feedpoint at its focus.
The ends of the mounting arms terminate in bolts that screw into a bracket to anchor the feedpoint.
The arms are mounted to the edges of the dish, so that they converge at a point along the axis of the dish.
The focus of the dish is not located at this point, but is located about 30 cm further from the dish.
We determined the focus acoustically in a quiet indoor location.
That is, the dish is large enough that a distinct change in sound quality can be heard when one's ear is in the vicinity of the focus.

It was necessary to manufacture a bracket that is compatible both with the feed assembly and
the mounting arms.
Figure \ref{fig:bracket} shows the mounting bracket used to hold the feed assembly with its
antenna at the focus of the dish.
Plans are located at \url{https://github.com/nick45508/MUL-2.4M-C-C-Band-Satellite-Dish-Technical-Schematics-and-Manual}.

The installation process subjects the bracket to substantial stresses, but the bracket is not
subjected to much stress once installed.
Therefore, the bracket was 3-D printed from Markforged’s Onyx filament, which is a micro
carbon fiber infused nylon filament with additional carbon fiber strands embedded within each
layer to add additional rigidity. Initially, the bracket was printed from Ultimaker Tough PLA,
however the PLA bracket cracked during use and the print flexed too much for continued use.
The carbon-fiber infused bracket solved these issues and produced a rigid part that has a
comparable tensile strength to aluminum, while being significantly lighter.

\subsubsection{Low noise amplifier feed assembly}

\begin{figure}[!htbp]
  \begin{center}
    \includegraphics[width=4in]{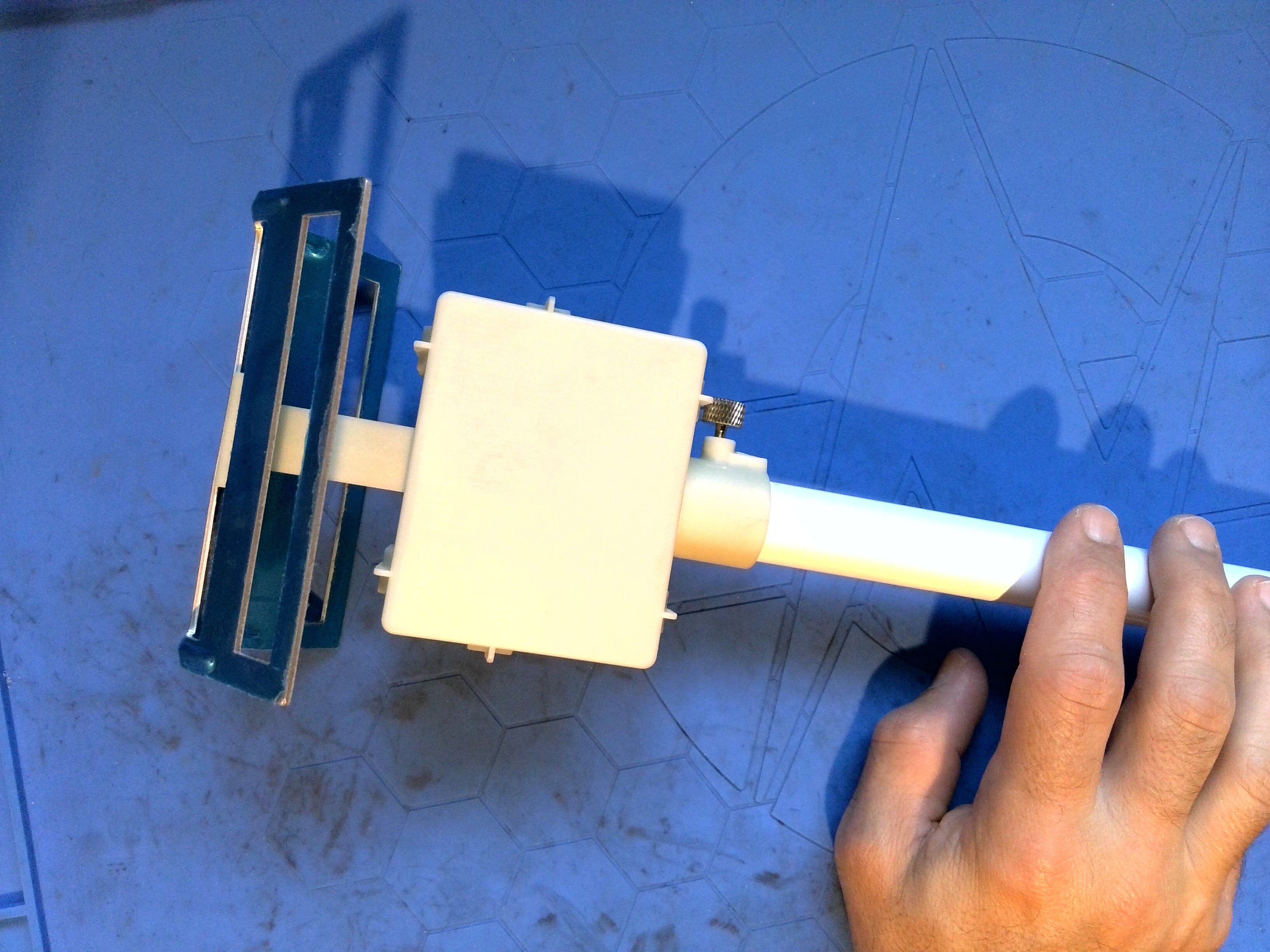}
    \caption{Feedpoint assembly}
    \label{fig:feed}
  \end{center}
\end{figure}

The feed assembly was purchased as a unit, and is attached to a pre-manufactured waterproof box:
S-band Discovery Feed: \url{https://www.crowdsupply.com/krakenrf/discovery-dish}.
Figure \ref{fig:feed} shows the structure of the feed assembly.
The feed assembly contains two integrated linearly polarized reflector antennas.
The Orion S-band transmitter is right hand circularly polarized (RHCP),
so using linear polarization amounts to about a $3$ dB loss.

Notice that the bracket (Figure \ref{fig:bracket}) and feed assembly (Figure \ref{fig:feed}) contain set screws which allows the feed assembly to slide and rotate along the axis of the dish.
This allows for finer adjustment of the feed assembly to correct for phase offsets present in the feed antenna.
To perform this fine adjustment, we located a strong signal from a GPS satellite and then maximized its signal strength.

\begin{figure}[!htbp]
  \begin{center}
    \includegraphics[width=4in]{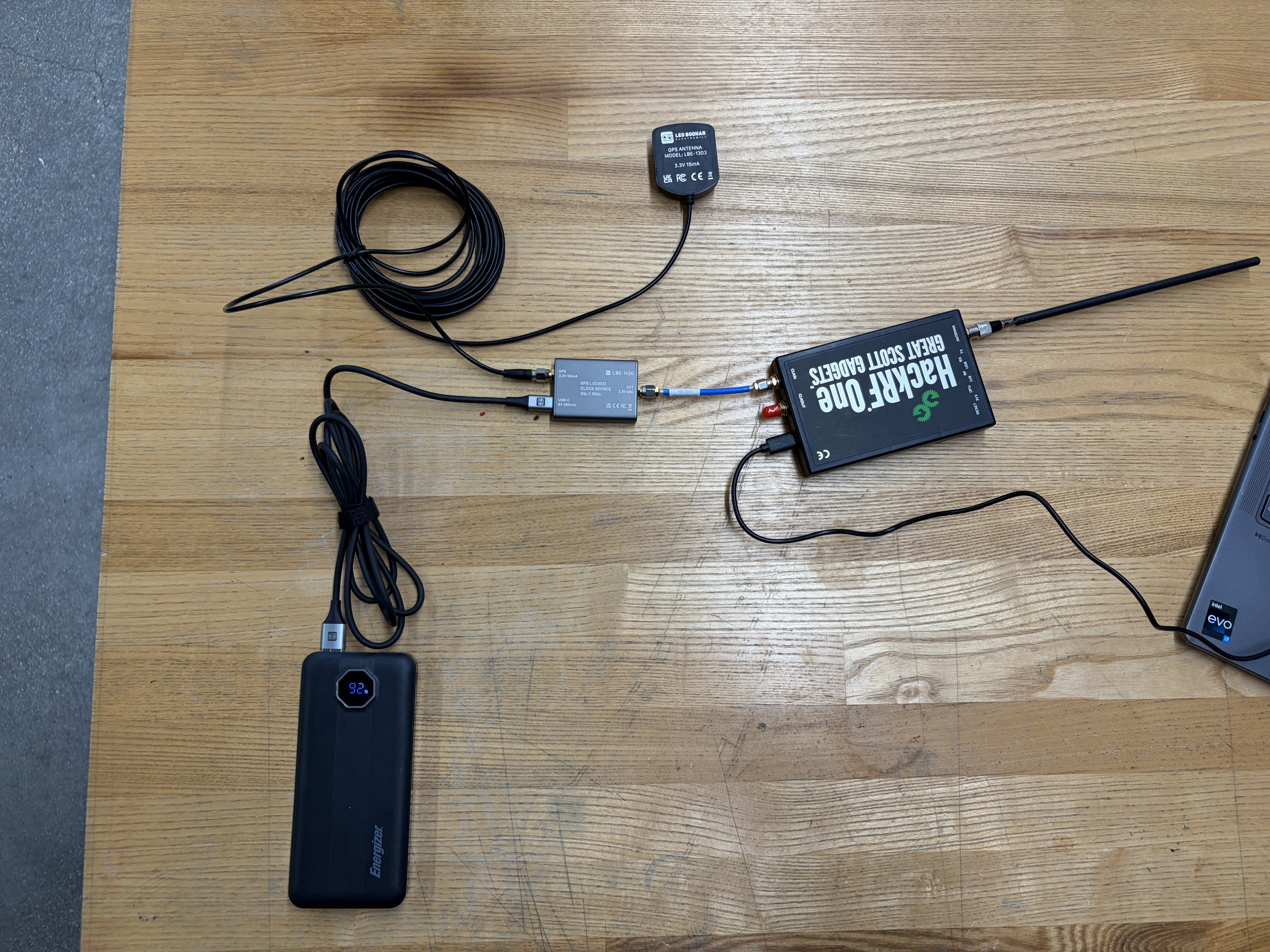}
    \caption{Detail of the RF chain; the rubber duck antenna is attached where the feed assembly would go during data collection.}
    \label{fig:rf_chain_detail}
  \end{center}
\end{figure}

The feed assembly is attached to the RF chain as shown in Figure \ref{fig:rf_chain_detail}.

\subsection{Radio collection chain}

\begin{figure}[!htbp]
  \begin{center}
    \includegraphics[width=4in]{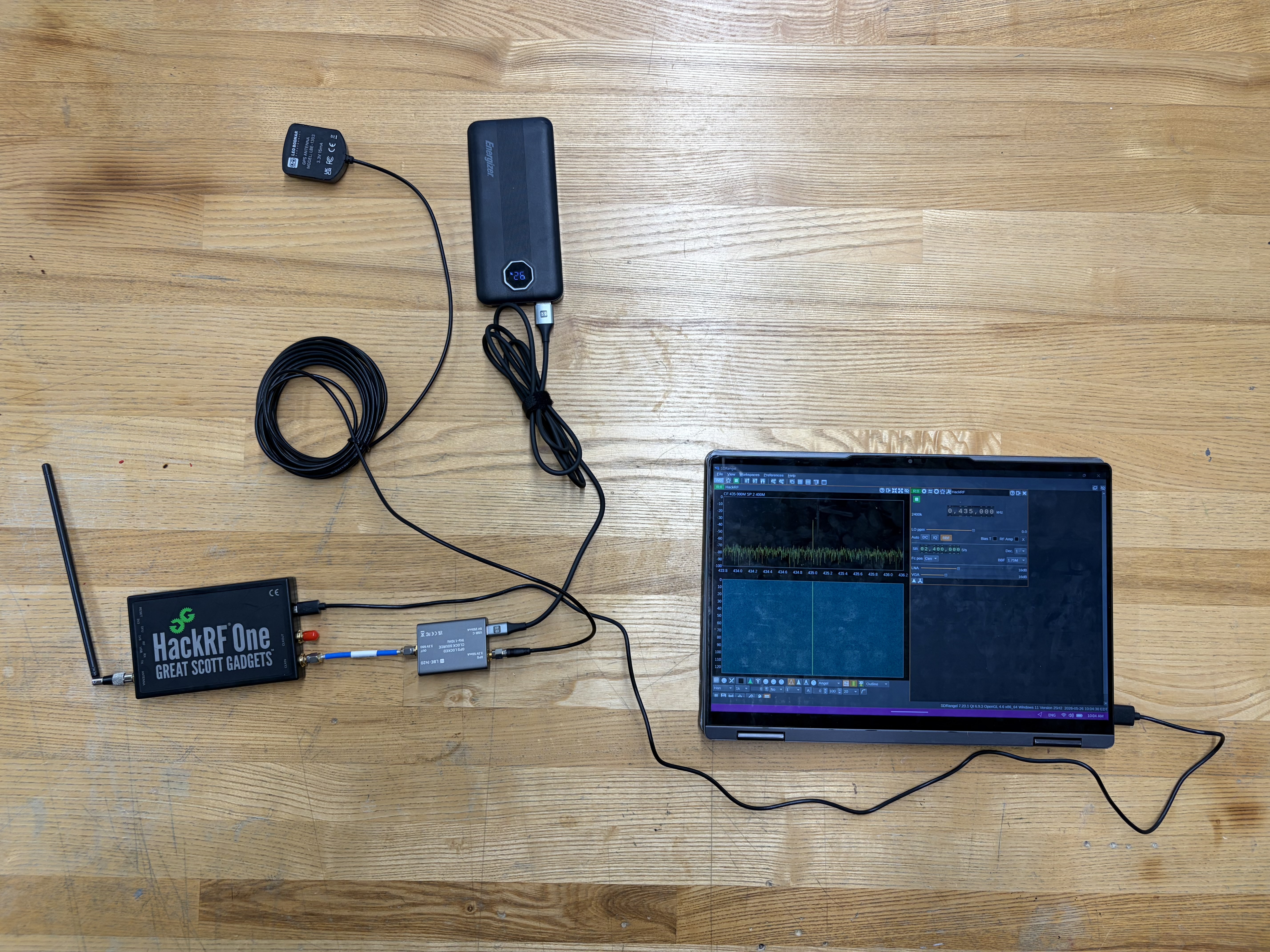}
    \caption{Overveiw of the RF chain.  The rubber duck antenna is attached where the feed assembly would go during data collection.  The computer shows data collection in progress.}
    \label{fig:rf_chain_overview}
  \end{center}
\end{figure}

\begin{figure}[!htbp]
  \begin{center}
    \includegraphics[width=4in]{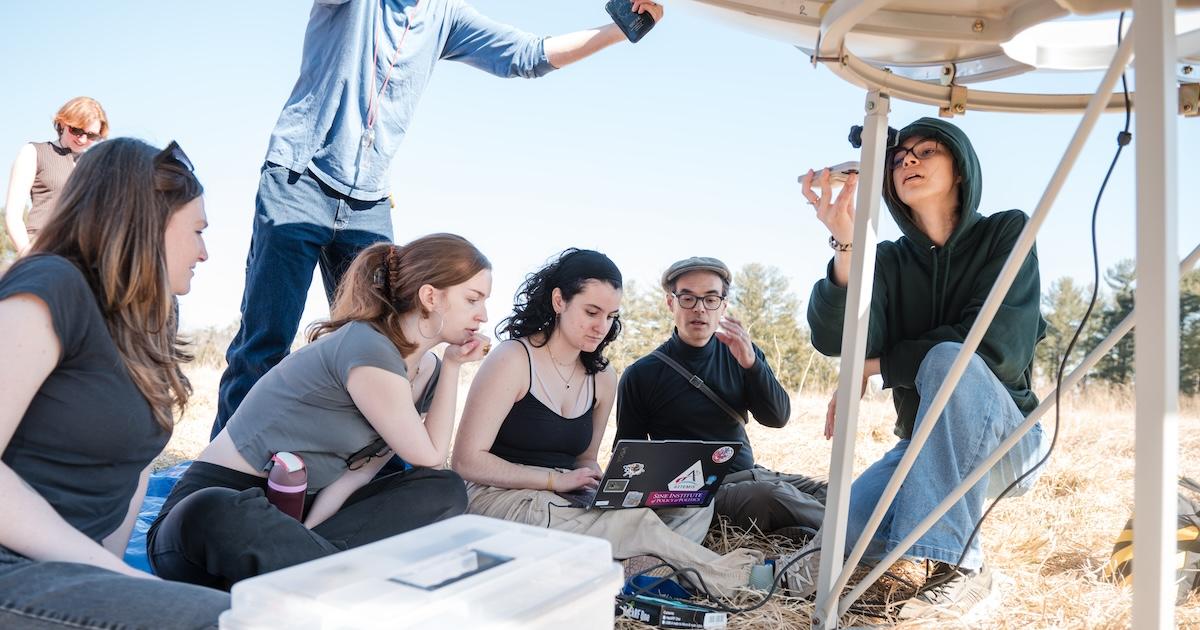}
    \caption{Data recording process.  (Photo by Nikolai Roster, CAS.)}
    \label{fig:pointing_and_recording}
  \end{center}
\end{figure}

Figure \ref{fig:rf_chain_overview} shows the entire RF collection chain as it would be assembled in the field, with a dummy rubber duck antenna in place of the actual feed assembly.
Each component shown in Figure \ref{fig:rf_chain_overview} corresponds to a block in Figure \ref{fig:feedline_schematic}.  
Figure \ref{fig:pointing_and_recording} shows the data collection in progress.

\subsubsection{HackRF software defined radio (SDR)}

The receiver used was a HackRF Software Defined Radio (SDR).
The receiver is capable of collecting 24 bit signed integer samples, at two per complex sample.
The sampling rate can be up to 20 MHz, though we used 5 MHz and 10 MHz.
The noise figure is about $10$ dB, as per measurements performed by others, \url{https://greatscottgadgets.com/2025/12-03-hackrf-pro-receive-sensitivity-and-noise-figure/}

\subsubsection{GPS disciplined clock}

To ensure an accurate time reference, we used a GPS Disciplined
Oscillator (GPSDO) in combination with the HackRF SDR to act as an
external clock for our observations. In this instance, the GPSDO
allows us to peg our timing reference off of Universal Standard Time
(UTC), ensuring an extremely accurate and reliable timing system. We
used the LBE-1420 GPSDO locked clock source running at 10 MHz, which
has a Allen deviation of $4.41\times 10^{-12}$ at $\tau = 100$ seconds and a phase deviation
of -147 dBc/Hz at a 1 kHz offset, both of which are more than
acceptable for our use case. The GPSDO is also a configuration
software that can be used alongside it to run diagnostics and change
the operating frequency, which can be found at

\url{https://www.leobodnar.com/shop/index.php?main_page=product_info&cPath=107&products_id=393}

\subsubsection{SDR Angel software}

\begin{figure}[!htbp]
  \begin{center}
    \includegraphics[width=4in]{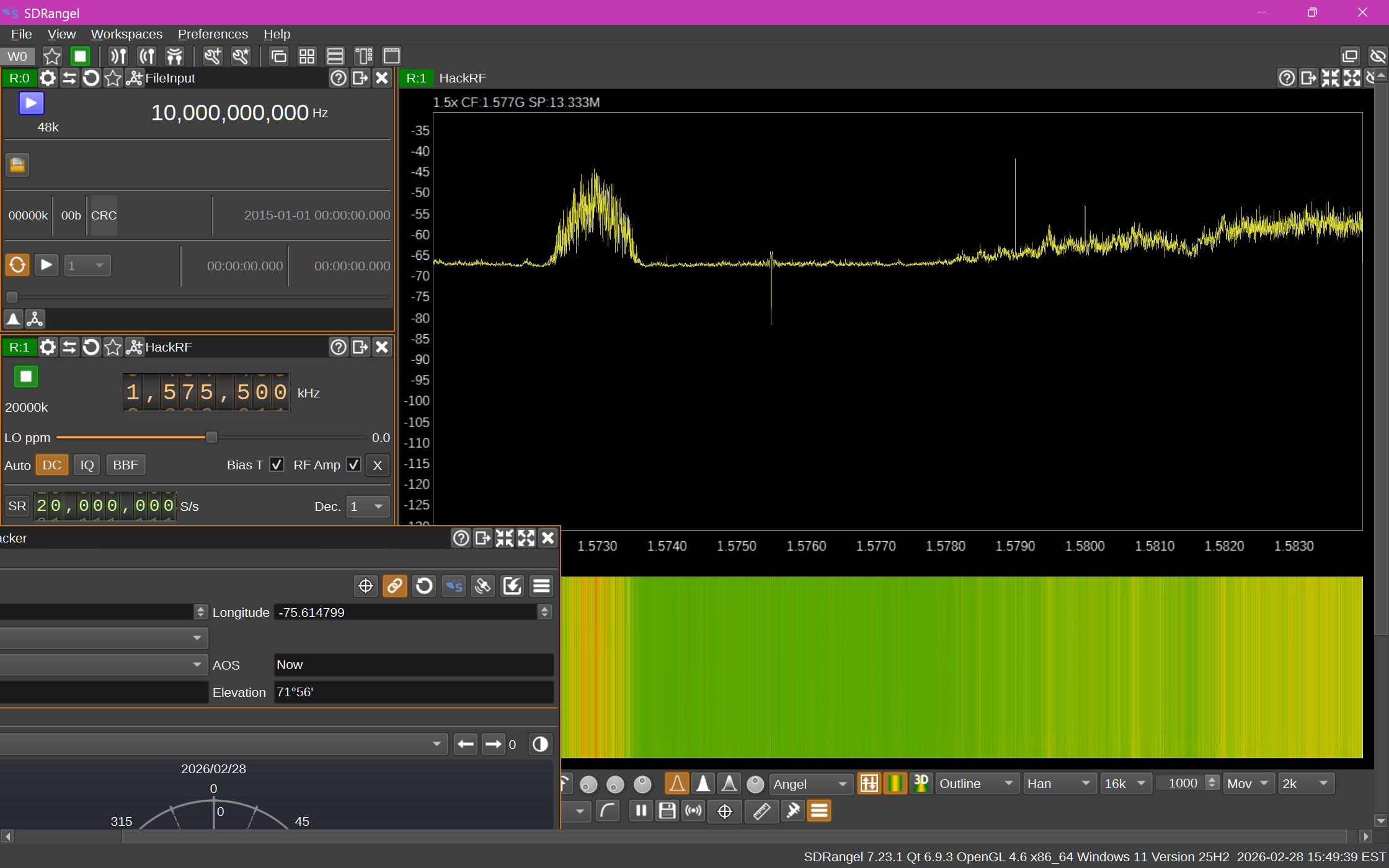}
    \caption{A typical screenshot of data collection in progress using SDR Angel}
    \label{fig:sdrangel}
  \end{center}
\end{figure}

We used the open source SDRAngel tool \url{https://www.sdrangel.org/} to control the SDR and for data collection.
SDRangel is an open-source software-defined radio (SDR) and signal analyzer frontend that supports a large range of hardware including most SDRs.  

After selecting the HackRF One as an input source, a file sink channel was added to record data. SDRangel can record data in either {\tt .wav} or {\tt .sdriq} file formats, with {\tt .wav} files being best suited for demodulated audio. The .sdriq format records in-phase/quadrature signals (I/Q), and best suited our needs. Files can also be imported into SDRangel for analysis. 
The {\tt .sdriq} file format is described in Section \ref{sec:file_format}.

\begin{figure}[!htbp]
  \begin{center}
    \includegraphics[width=4in]{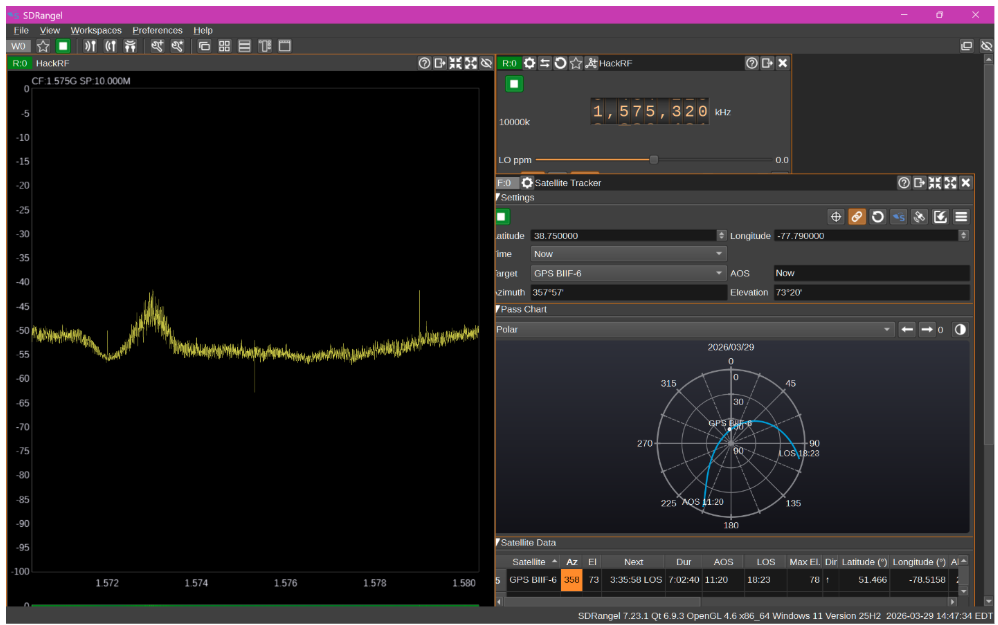}
    \caption{SDR Angel used for satellite tracking.}
    \label{fig:sdrangel_tracker}
  \end{center}
\end{figure}

Although it was not used to track Artemis II, SDRangel's satellite tracking feature can be used to download and display a table of data from two line elements (TLEs) based on the user's position that provides useful information such as acquisition of signal (AOS), loss of signal (LOS), azimuth, elevation, time until and the duration of the next pass, as well as polar and cartesian path plots for a large variety of satellites. Figure \ref{fig:sdrangel_tracker} shows the feature being used to track GPS BIIF-6. 

\subsection{Doppler processor}

The Doppler processor is written in Python, using NumPy, SciPy, and PyTorch.
It can optionally support the use of a GPU for acceleration.
The source code for the Doppler processor is located at \url{https://github.com/kb1dds/artemis_sdriq}

If $r=r(t)$ is range from the observer to the Orion spacecraft, which is transmitting at a center frequency of $f_0$, then the difference between the transmitter center frequency and the received frequency is the \emph{Doppler frequency},
\begin{equation*}
  \Delta f =  f_0 \frac{2}{c} \frac{dr}{dt},
\end{equation*}
where $c$ is the speed of light.

\subsubsection{File format}
\label{sec:file_format}

SDR Angel records I/Q samples in its own SDRIQ format.
The format is described on GitHub \url{https://github.com/f4exb/sdrangel/blob/master/plugins/samplesource/fileinput/readme.md}.
Each file contains a 32 byte header, shown in Table \ref{tab:sdriq_header}.
After the header, the samples follow, with alternating real and complex signed integers.
If the header specifies 24 bit samples, the samples are aligned to 32 bit boundaries.
A constant tone at the center frequency of the file is represented as an alternating sequence $\pm 1$.

\begin{table}
  \begin{center}
    \caption{SDRIQ file header contents}
    \label{tab:sdriq_header}
    \begin{tabular}{|l|c|l|l|}
      \hline
      Offset & Length & Numpy type & Contents \\
      (bytes) & (bytes) & &\\ 
      \hline
      0 & 4 & {\tt uint32} & Sample rate in samples/second\\
      4 & 8 & {\tt uint64} & Center frequency in Hz \\
      12 & 8 & {\tt uint64} & Unix timestamp of start in milliseconds \\
      20 & 4 & {\tt uint32} & Sample size in bits \\
      24 & 4 & {\tt uint32} & Padding with zeros \\
      28 & 4 & {\tt uint32} & CRC of previous 28 bytes (ignored)\\
      \hline
    \end{tabular}
  \end{center}
\end{table}

\subsubsection{Nonlinear detector via $z^4$ process}

The Orion spacecraft's S-band transmitter is not directly intended for ranging and Doppler estimation.
Instead, it utilizes a quadrature phase shift keyed (QPSK) signal format.
The message to be transmitted is sent as a sequence utilizing four possible symbols,
realized as four possible phase shifts applied to a sinewave carrier.
These phase shifts are most conveniently expressed as the complex numbers $\{1, i, -1, -i\}$.
The message is therefore a function $m=m(t)$, where $m(t)$ is one of those four possible complex numbers.
(Note: the Orion S-band signal is actually somewhat more complicated, because \emph{pulse shaping} is applied to $m$ to control its bandwidth.  We ignored that effect in our Doppler processor.)

Given a center frequency of $f_0$, the received QPSK signal will be of the form
\begin{equation*}
  s(t) = m(t) e^{2\pi i (f_0 + \Delta f) t + \phi},
\end{equation*}
where $\phi$ is an unknown phase offset.

Our goal is to determine $\Delta f$ from $s(t)$.
If $m(t)$ were not present, $\Delta f$ could be estimated by examining the spectrogram of $s(t)$ for peaks.
The signal will have a spectral peak at $f_0+\Delta f$, and since $f_0$ is known, the $\Delta f$ can be easily obtained.

The presence of the message $m(t)$ tends to spread the spectrum across a bandwidth of roughly that of the symbol rate of $m(t)$,
which reduces the SNR of the spectral peak at $f_0+\Delta f$.
We aim to accumulate the signal back into a single spectral bin.
Notice that raising $m(t)$ to the fourth complex power, namely $(m(t))^4$, results in a constant function, because all of the possible symbols are roots of unity.  This process can be applied to the entire signal, which effectively removes the message from the signal but retains the Doppler.

The cleanest process is to \emph{baseband} the signal first, which means we remove the effect of the carrier $f_0$ by modulation,
\begin{equation*}
  b(t) := s(t) e^{-2\pi i f_0 t} =  m(t) e^{2\pi i \Delta f t + \phi}.
\end{equation*}
This allows us to apply a fixed lowpass filter to remove noise away from the carrier.
We then raise the resulting signal to the fourth power,
\begin{equation*}
  (b(t))^4 = \left(s(t)e^{-2\pi i f_0 t} \right)^4 = e^{2\pi i (4 \Delta f) t + 4 \phi}.
\end{equation*}
The result of this process now has a single spectral peak at $4$ times the Doppler frequency.

\subsubsection{Doppler sweep}
\label{sec:sweep}

Since the Doppler frequency may take any real value,
it is possible that the spectral peak from the Doppler frequency is quite narrow.
As such, the Doppler spectral peak can be missed due to aliasing between frequency bins if we use a discrete Fourier transform.
Therefore, we employ a modified process wherein we baseband using a candidate Doppler $\widehat{\Delta f}$, namely
\begin{equation*}
  \left(s(t)e^{-2\pi i (f_0 + \widehat{\Delta f}) t} \right)^4 = e^{2\pi i \left(4 \left(\Delta f-\widehat{\Delta f}]\right)\right) t + 4 \phi}.
\end{equation*}
Sweeping over possible values of $\widehat{\Delta f}$, we obtain a peak when $\widehat{\Delta f} = \Delta f$.
This gives somewhat finer control over the process at the expense of greater computation time.
Since this is effectively a zero frequency peak, the Doppler processor can be understood as detecting the peaks of the function
\begin{equation*}
  w\left(\widehat{\Delta f}\right) := \sum_t \left(s(t)e^{-2\pi i (f_0 + \widehat{\Delta f}) t} \right)^4.
\end{equation*}
Note also that the detected peak in this case does not have the factor of $4$ applied to the detected Doppler frequency.

\begin{figure}[!htbp]
  \begin{center}
    \includegraphics[width=5in]{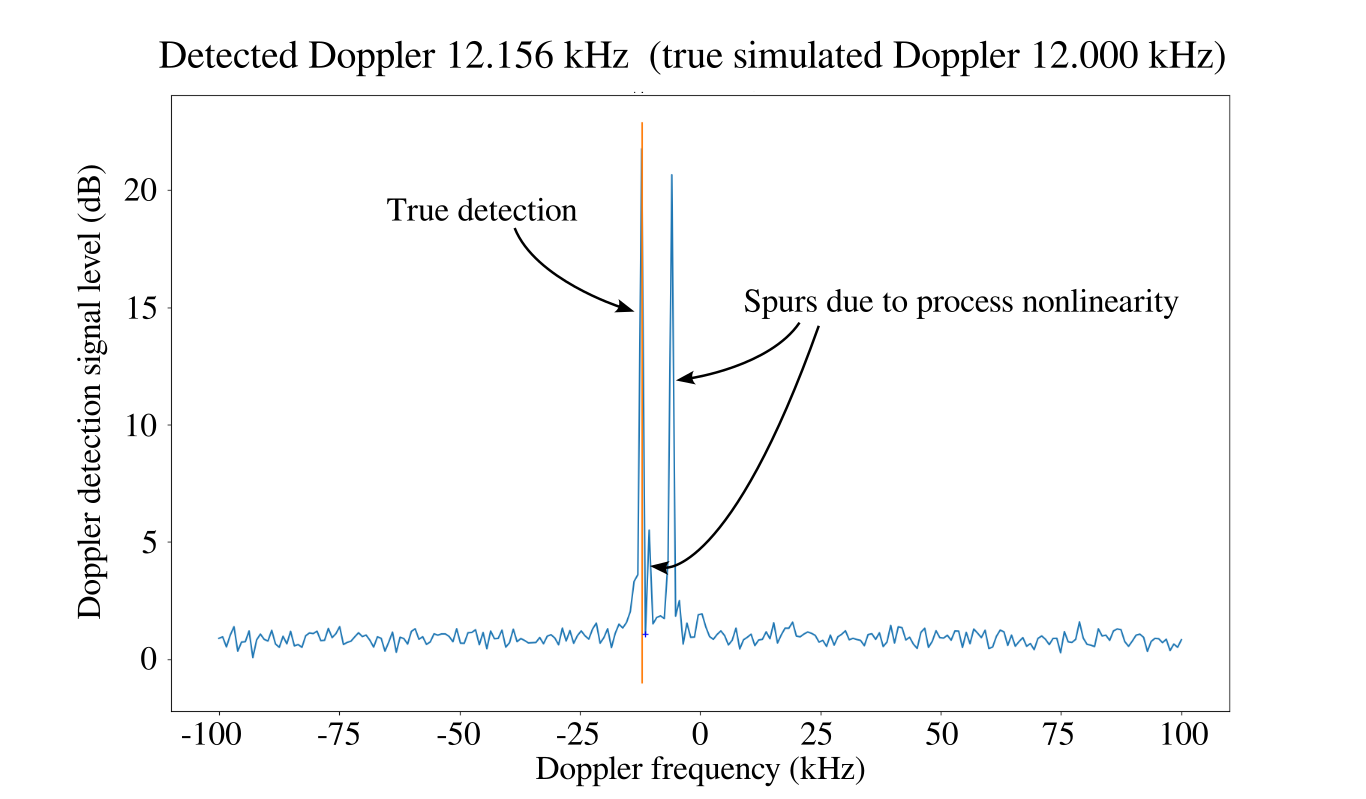}
    \caption{Simulated Doppler detection sweep.  True Doppler at 12 kHz is detected, though ambiguity spurs are present at sub-harmonics.}
    \label{fig:doppler_example_w_nonlinearity}
  \end{center}
\end{figure}

Figure \ref{fig:doppler_example_w_nonlinearity} shows a typical sweep of $\widehat{\Delta f}$.
Note that the nonlinearity of the fourth power creates additional spurs at sub-harmonics with low SNR.
Some signal loss is present, and in some cases the sub-harmonic spurs can have higher SNR than the true Doppler.

\subsubsection{GPU batching}

Because the signal is anticipated to be very weak,
large processing windows can provide gain.
Because the operations involved are highly parallelizable,
we used an NVIDA GeForce RTX 5070 GPU for the processing.
Using the GPU offers roughly a factor of $5\times$ speedup over CPU processing on the same machine.

There are three main sample size parameters used in the Doppler processor:
\begin{enumerate}
\item The processing window for basebanding and filtering (the length of $b(t)$),
\item The sweep window (number of candidate $\widehat{\Delta f}$ values tested), and
\item The number of consecutive windows to average the sweeps (number of copies of $w(\widehat{\Delta f})$).
\end{enumerate}
Processing window, sweep, and averaging sizes are all user-specifiable,
and can be adjusted to fit the resulting job into the available memory.

For basebanding, we used processing windows of 65536 samples long,
ran the Doppler sweep across 512 candidate Doppler frequencies,
and averaged these over 7680 consecutive windows.
The memory requirements to store this entire set of samples in RAM are larger than the VRAM for the GPU we used,
so the job was split into 15 batches.
We also explored larger window sizes (up to 262144 samples)
and larger numbers of consecutive windows to average (up to the use of each entire collected data file).
These larger windows did not seem to improve results beyond what is shown in this report.

\section{System validation}

We performed several forms of validation:
\begin{enumerate}
  \item CAMRAS archival data from Artemis I,
  \item Two collections of our own: a cubesat of opportunity, a GPS satellite, and
  \item Simulation validation for doppler processor.
\end{enumerate}

\subsection{Trajectory}

As per NASA's instruction, JPL Horizons \url{https://ssd.jpl.nasa.gov/horizons/app.html} is used as the sole ephemeris source for the Artemis II mission.
We validated our location and time inputs by using several different nearby ground locations (``Washington, DC'' text input versus direct lat/lon for Airlie Farm),
and had several team members perform independent queries.

We used gpredict \url{https://github.com/csete/gpredict} for the practice runs collecting LEO and GEO satellites.

\subsection{Radio collection}

\subsubsection{Cubesat of opportunity}

During a practice run the weekend before the Artemis II launch, we collected a passing cubesat at 2026-03-29T18:57:03.024.

\begin{figure}[!htbp]
  \begin{center}
    \includegraphics[width=4in]{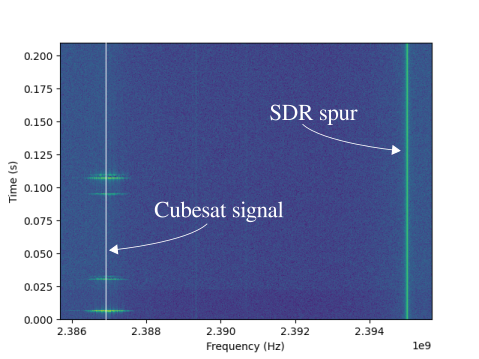}
    \caption{Spectrogram of the cubesat of opportunity.}
    \label{fig:cubesat_spectrogram}
  \end{center}
\end{figure}

According to the spectrogram shown in Figure \ref{fig:cubesat_spectrogram}, the center frequency is $2.387$ GHz.
This frequency is outside amateur frequencies.
Page 91 of \url{https://www.fcc.gov/sites/default/files/fcctable.pdf} indicates that the center frequency is located in a band allocated for ``Space Operation'',
so it is probably an industry- or government-operated satellite.
It could be a Starlink control frequency, but these are not published.

\subsubsection{GPS satellite}

After collecting data from the Artemis II spacecraft but before packing up to return to campus,
we successfully collected a GPS satellite at 2026-04-03T07:30. 

\begin{figure}[!htbp]
  \begin{center}
    \includegraphics[width=3in]{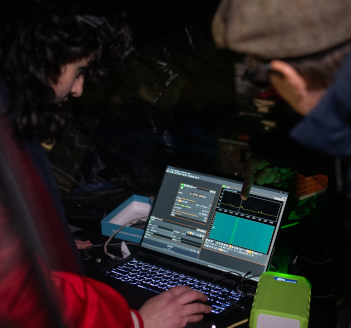}
    \caption{Reception of a GPS satellite.  (photo by Nikolai Roster, CAS)}
    \label{fig:gps_reception}
  \end{center}
\end{figure}

The signal was visually confirmed at the correct frequency for GPS signals, shown in Figure \ref{fig:gps_reception}.
No data were recorded from this satellite.

\subsection{Dynamic range of SDR}

The SDR produces 24 bit samples as signed integers.
One bit is sign, so only 23 bits for mantissa.
This corresponds to a dynamic range of
\begin{equation*}
  2^{23} = 8 \times 10^{6} = 10^{6.9} =10^{138/20} = 138 \text{ dB}.
\end{equation*}

\subsection{SDR artifacts}

\begin{figure}[!htbp]
  \begin{center}
    \includegraphics[width=4in]{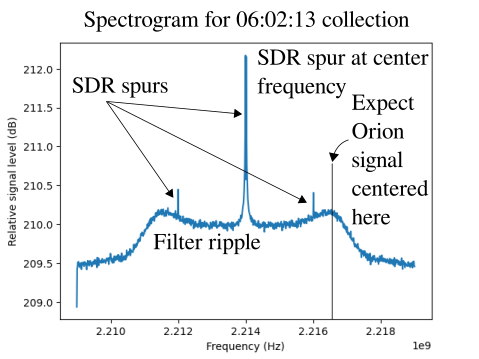}
    \caption{Typical spectrogram collected in the Artemis II frequency band, showing SDR spur artifacts.}
    \label{fig:typical_spectrogram}
  \end{center}
\end{figure}

Figure \ref{fig:typical_spectrogram} shows a snapshot of the spectrogram of some of our data,
calling out spurs every 1 MHz on the graphic.

\subsection{Doppler processing}

To validate the performance of our Doppler processor,
we performed two tests:
\begin{enumerate}
\item A simulation study wherein a known QPSK signal was injected into background noise with known SNR, and
\item Processing of archival data collected from Artemis I.
\end{enumerate}

\subsubsection{Simulation study}
\label{sec:doppler_sim_validation}

The simulation study injected a QPSK signal with the same symbol rate as the Orion S-band transmitter (4 MHz) with a random-but-known message into a background of additive white Gaussian noise.
The SNR of the signal was varied from -20 dB to 5 dB,
and the processing window size was varied by choosing powers of 2 from 256 to 8192.
The strength of the Doppler peak (of $w(\widehat{\Delta f})$ in Section \ref{sec:sweep}) was recorded for each.
A new noise signal was generated for each scenario,
but the signal was left unchanged.

\begin{figure}[!htbp]
  \begin{center}
    \includegraphics[width=4in]{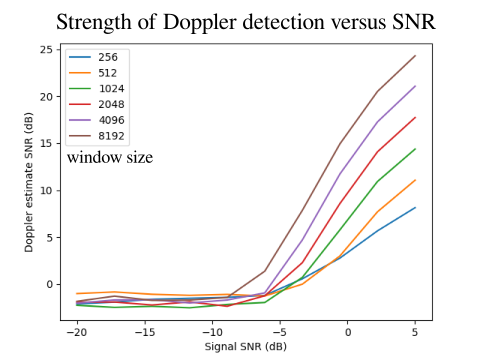}
    \caption{Performance of the Doppler processor as a function of SNR.  Processing window sizes are shown as separate curves.}
    \label{fig:doppler_perfvalidation}
  \end{center}
\end{figure}

Figure \ref{fig:doppler_perfvalidation} shows the strength of the Doppler signal peak versus the SNR of a simulated QPSK signal with symbol rate 4 MHz.  The main conclusion from this experiment is that the Doppler processor gives
\begin{itemize}
\item Reliable results if SNR $\ge -5$ dB,
\item Possible results if SNR $\ge -10$ dB and a large window size is used, but is
\item Unlikely to succeed if SNR $\le -10$ dB.
\end{itemize}
Notice that in Section \ref{sec:link_budget} the best case SNR for our collection window is $-15$ dB.
It is therefore unlikely that we will be able to measure a Doppler frequency with a window size smaller that $8192$ samples.

\subsubsection{Using Artemis I archival data}

The Dwingeloo observatory at $52.8^\circ$ N, $6.38^\circ$ E collected data from the Artemis I mission using their 25 m parabolic dish.
These observations are archived generally at \url{https://data.camras.nl/artemis/}.

We considered the collection with timestamp 2022-11-30T19:17:04,
which was recored with a 5 MHz sampling rate, centered at 2.2165 GHz.
The specific link to this collection is \url{https://data.camras.nl/artemis1/?datafile=unflaggedresiduals_quad_20221130.txt}.
At this time, the spacecraft was in orbit around the moon.
JPL Horizons indicates that the instantaneous range rate from Dwingeloo was about $-0.11$ km/s.

\begin{figure}[!htbp]
  \begin{center}
    \includegraphics[width=4in]{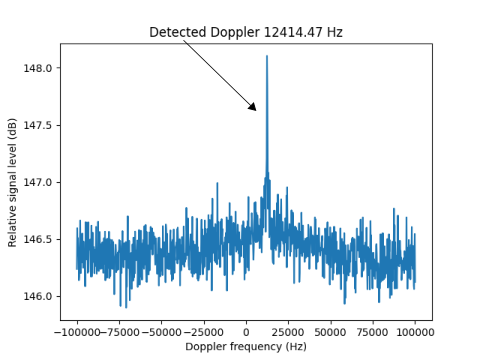}
    \caption{Doppler sweep of archival Artemis I data showing detected Doppler frequency of 12 kHz.}
    \label{fig:artemis1_dopplerscan}
  \end{center}
\end{figure}

As Figure \ref{fig:artemis1_dopplerscan} shows, our Doppler processor finds a strong doppler signal at $12.4$ kHz.
The Dwingeloo observatory itself reports a Doppler of $24.2$ kHz at this time.
Given the fact that our nonlinear process may detect ambiguous sub-harmonics instead of the true Doppler when the SNR is low
(see Figure \ref{fig:doppler_example_w_nonlinearity}),
it is possible that our estimate and Dwingeloo's estimate are consistent.
The Doppler we detected corresponds to a range rate of
\begin{equation*}
  \frac{1}{2} 1.2 \times 10^{4} \text{Hz} / 2.2165 \times 10^{9} \text{Hz} \times 3 \times 10^{8} \text{m/s} = 0.8 \text{ km/s}.
\end{equation*}

\subsubsection{Cubesat of opportunity}

\begin{figure}[!htbp]
  \begin{center}
    \includegraphics[width=4in]{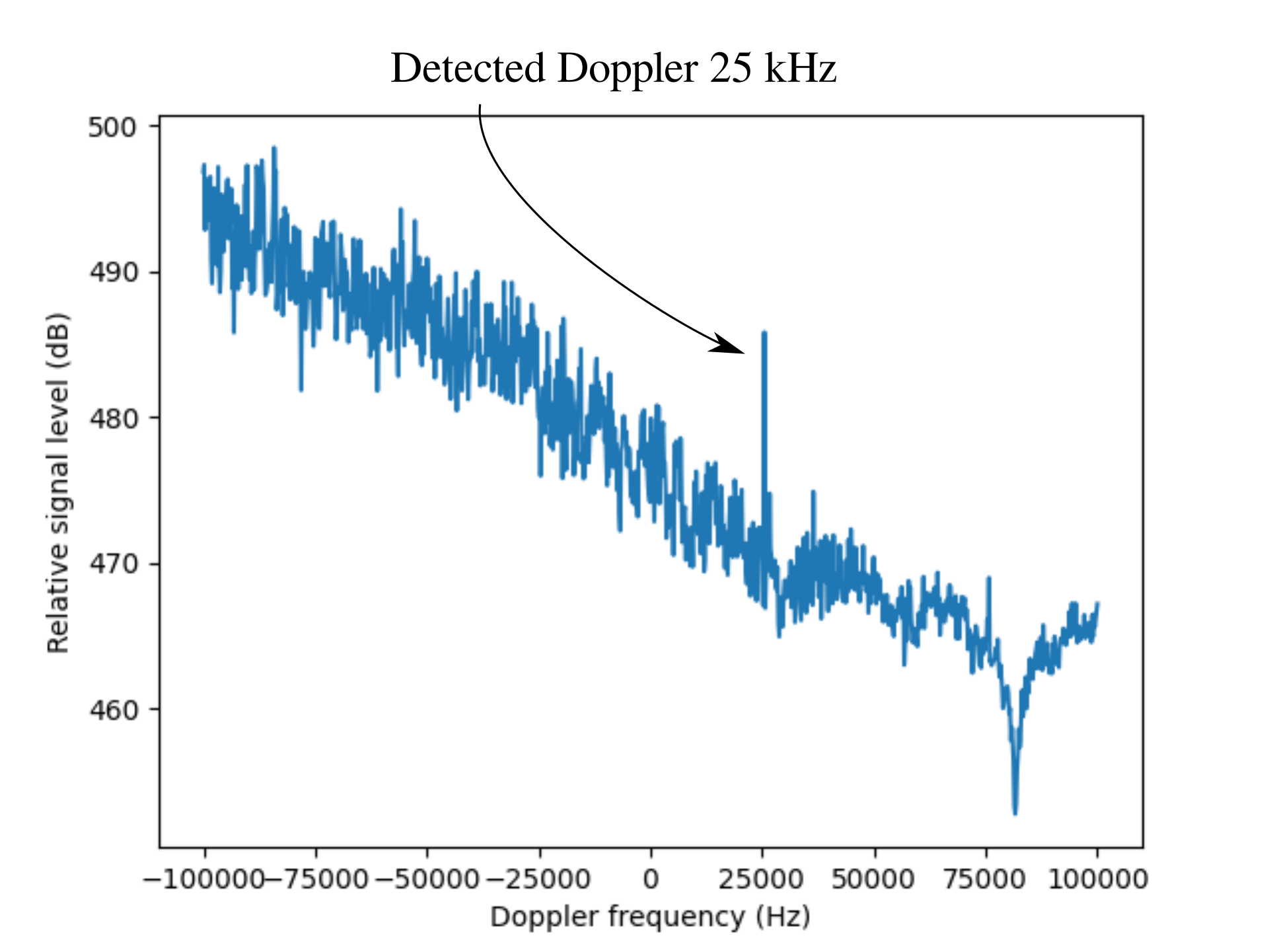}
    \caption{Doppler sweep of cubesat of opportunity showing detected Doppler frequency of 25 kHz.}
    \label{fig:cubesat_dopplerscan}
  \end{center}
\end{figure}

Our Doppler processor might be able to determine a Doppler frequency for the cubesat of opportunity,
assuming that its signal consists of bursts of phase shift keyed (PSK) signals, if the number of phases used is divisible by $2$.
Figure \ref{fig:cubesat_dopplerscan} shows that we obtained a strong Doppler peak at about $25$ kHz from the cubesat,
This corresponds to
\begin{equation*}
  \frac{1}{2} 2.5 \times 10^{4} \text{Hz} / 2.387 \times 10^{9} \text{Hz} \times 3 \times 10^{8} \text{m/s} = 1.5 \text{ km/s},
\end{equation*}
which is a reasonable relative velocity for a LEO satellite.

\section{Experimental results}

\begin{figure}[!htbp]
  \begin{center}
    \includegraphics[width=4in]{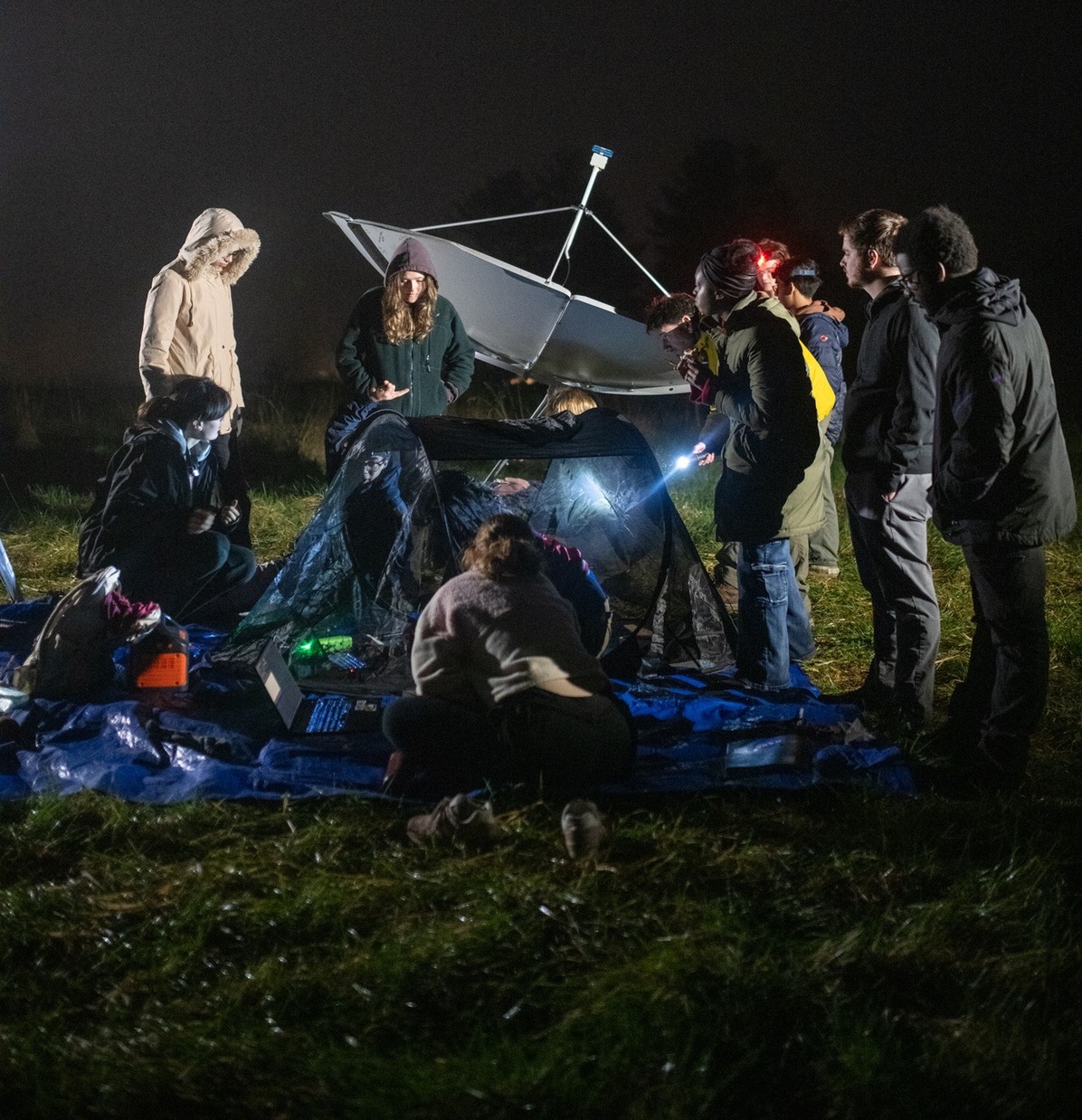}
    \caption{Collecting data from a GPS satellite.  (Photo by Nikolai Roster, CAS.)}
    \label{fig:collection_in_progress}
  \end{center}
\end{figure}

An intrepid group of students (see Figure \ref{fig:collection_in_progress}) made the collection attempt from 1-4 am local time early Friday morning (3 April 2026).

\subsection{Trajectory considerations}

We had hoped that from our observing site that the Orion spacecraft would be above the horizon shortly after the trans-lunar injection (TLI) burn,
as this would give us enough SNR to perform Doppler measurements.
We based our initial plan on a preprint paper \cite{aas_25_100} which included a planned timeline.

Unfortunately, our first observing window opened after the spacecraft had passed the geostationary belt (45000 km) after TLI.
Additionally, JPL Horizons indicated that the next observing window within 100000 km would be on the same day as the splashdown,
and then the spacecraft was at least $10^\circ$ above the horizon (clearing the trees at our observing site) for a narrow window of a few hours in the early morning.
Because the students were also hosting a ``splashdown party'' later that evening,
we opted to attempt data collection during the outbound window only.
For our chosen window, the derived pointing information from JPL Horizons is summarized in Table \ref{tab:pointing}.

\begin{table}
  \begin{center}
    \caption{JPL Horizons-derived pointing data for Artemis II from Airlie Farm}
    \label{tab:pointing}
    \begin{tabular}{|l|l|l|l|l|l|}
      \hline
      UTC time & Az & El & Range & Range rate & Doppler\\
      & (True N) & &  &  & \\
      \hline
      2026-04-03T05:00 &$145^\circ$ & $18^\circ$ & 74000 km & 2.65 km/s & 39.2 kHz\\
      2026-04-03T06:00 &$158^\circ$ & $22^\circ$ & 83000 km & 2.52 km/s & 37.2 kHz \\
      2026-04-03T07:00 &$171^\circ$ & $25^\circ$ & 92000 km & 2.43 km/s & 36.0 kHz \\
      2026-04-03T08:00 &$185^\circ$ & $25^\circ$ & 101000 km & 2.38 km/s & 35.2 kHz \\
      \hline
    \end{tabular}
  \end{center}
\end{table}

Based upon the pointing information, our Doppler sweeps should therefore generally consider 25--45 kHz.

\subsection{Link budget from trajectory}
\label{sec:link_budget}

The worst case transmitter power is stated as $39$ dBm.
For our observation window (see Table \ref{tab:pointing}), the worst case range is about $10^5$ km.
The path loss at this range is about $20 \log_{10}(10^5)+20\log_{10}(2.2615)+92.45 =200$ dB.
The antenna gain is $32$ dB.
Polarization mismatch loss is about $3$ dB.
The signal at the antenna feedpoint is then $-135$ dBm.

Using $k= -228$ dBW/K/Hz, at $290$ Kelvin operating temerature,
\begin{equation*}
  kT = (-228 + 10\log_{10} 290 + 30) = -173 \text{ dBm/Hz}.
\end{equation*}
If the signal has $4$ MHz bandwidth, the best case noise floor is
$kTB = -173 + 66 = -107$ dBm.
Hence the SNR is $-25$ dB.

If transmitter power is the higher of the two options, namely $49$ dBm,
the SNR is $-15$ dB,
which is tantalizingly close given the Doppler performance analysis shown in Section \ref{sec:doppler_sim_validation}.
We are a ``go'' for the collection attempt!

\subsection{Data products collected}
\label{sec:data_summary}

We collected $6$ separate recordings of sufficient size to warrant Doppler processing,
shown in Table \ref{tab:data_collections}.

\begin{table}
  \begin{center}
    \caption{Data products collected for Artemis II}
    \label{tab:data_collections}
    \begin{tabular}{|l|l|l|l|}
  \hline
  UTC time & Length & Center & Samplerate\\
  \hline
  2026-04-03T06:02:13.071 & 199.85 s & 2.2140 GHz & 10 MHz \\
  2026-04-03T06:11:19.705 & 396.20 s & 2.2140 GHz & 10 MHz \\
  2026-04-03T06:32:00.823 & 144.88 s & 2.2140 GHz & 10 MHz \\
  2026-04-03T06:36:21.044 & 439.81 s & 2.2165 GHz & 5 MHz \\
  2026-04-03T06:51:25.639 & 605.75 s & 2.2165 GHz & 5 MHz \\
  2026-04-03T07:10:42.219 & 402.00 s & 2.2165 GHz & 5 MHz \\
  \hline
\end{tabular}
\end{center}
\end{table}

The observed noise floor is $-90$ dB in all collections,
which is notably higher than our estimate of $-107$ dB.
This means that estimated SNR is between $-129 + 90 = -39$ dB and $-29$ dB.
This should be within the dynamic range of the receiver.
But even the best case SNR is considerably less than required for the Doppler processor (Section \ref{sec:doppler_sim_validation}),
so successful Doppler measurement is unlikely.

\begin{figure}[!htbp]
  \begin{center}
    \includegraphics[width=4in]{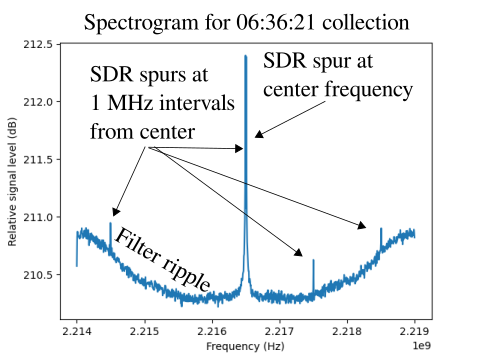}
    \caption{Spectrogram of the 06:36:21 dataset.}
    \label{fig:06_36_21_spectrogram}
  \end{center}
\end{figure}

Figure \ref{fig:06_36_21_spectrogram} shows a typical spectrogram from our data.
This spectrogram shows several visible SDR spurs (at 1 MHz intervals from the center frequency),
but no visible signal from the Orion spacecraft.

\subsection{Doppler processing summary}

\begin{figure}[!htbp]
  \begin{center}
    \includegraphics[width=4in]{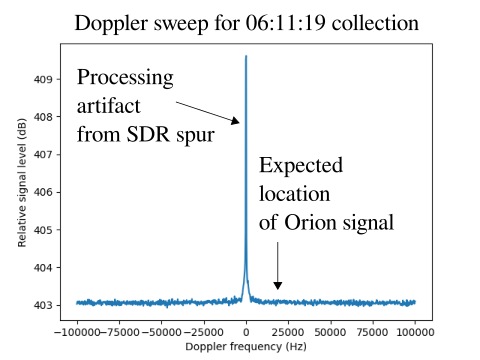}
    \caption{Doppler scan of the 06:11:19 dataset.  No visible significant peaks are present beyond an artifact at 0 Hz caused by the SDR spur.}
    \label{fig:06_11_19_dopplerscan}
  \end{center}
\end{figure}

Figure \ref{fig:06_11_19_dopplerscan} shows a 200 kHz Doppler sweep for the 06:11:19 collection,
which unfortunately---but not surprisingly---does not show any clear Doppler signature.

\begin{figure}[!htbp]
  \begin{center}
    \begin{tabular}{ccc}
      \includegraphics[width=2in]{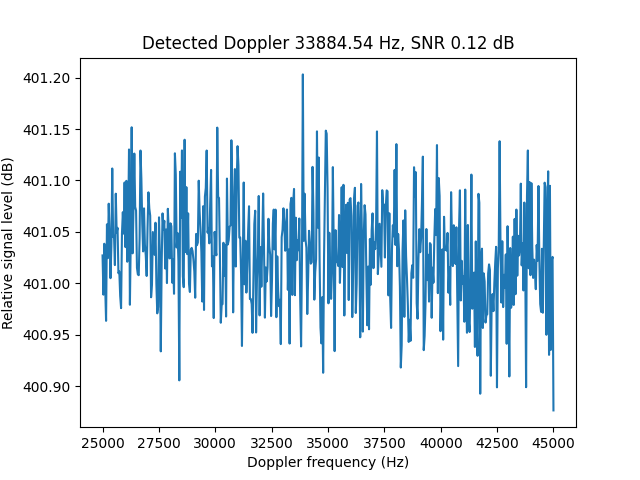}&
      \includegraphics[width=2in]{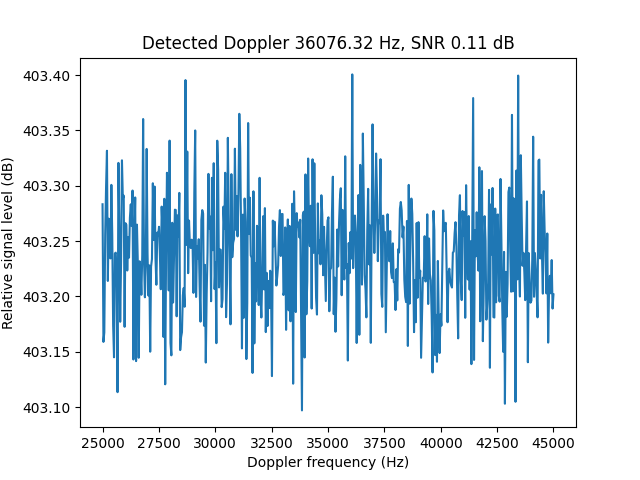}&
      \includegraphics[width=2in]{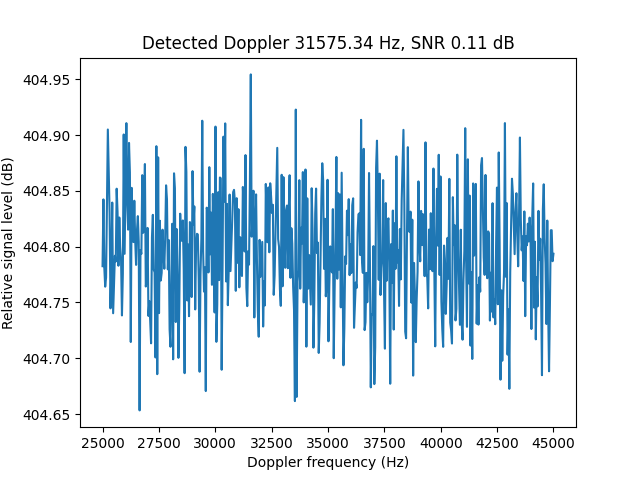}\\
      \includegraphics[width=2in]{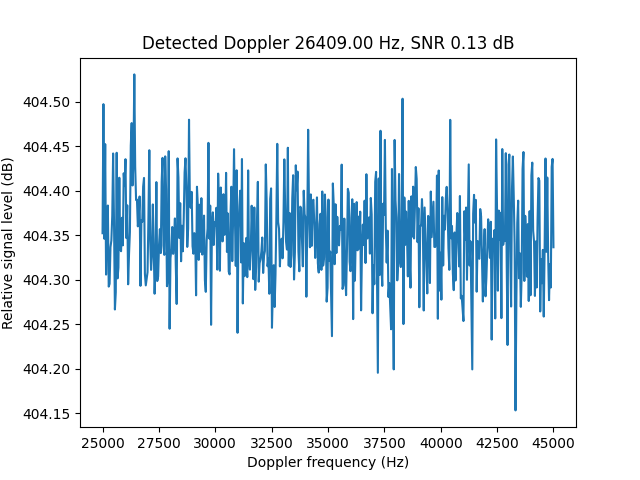}&
      \includegraphics[width=2in]{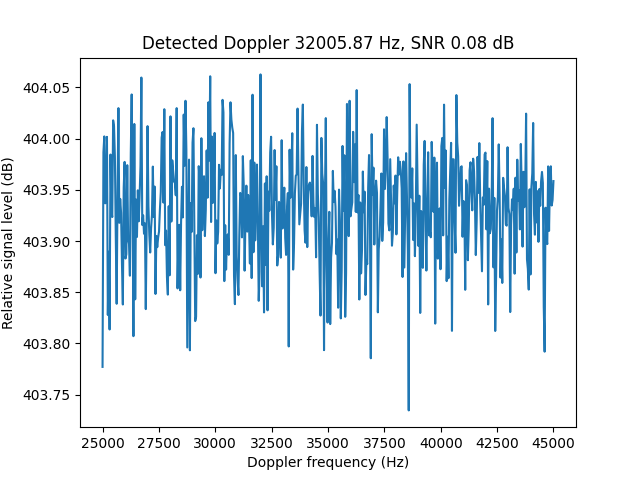}&
      \includegraphics[width=2in]{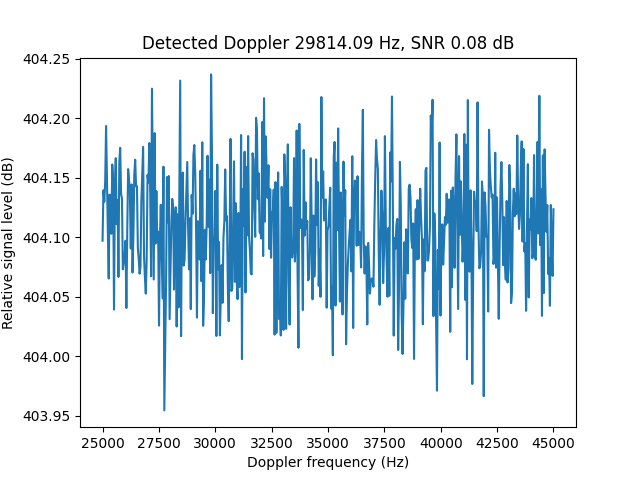}\\
    \end{tabular}
    \caption{Doppler scans of all attempted Artemis II collections using 25-45 kHz sweep.}
    \label{fig:dopplerscans}
  \end{center}
\end{figure}

We processed all files in batch using the following parameters; total runtime was about 25 hours,
\begin{enumerate}
\item Processing window size 65536 samples,
\item Doppler sweep 25--45 kHz in 512 steps,
\item Averaging 512 consecutive windows in batches of 15, and
\item No baseband filtering applied.
\end{enumerate}
We also tried applying baseband filtering of 4 MHz and 6 MHz,
and a processing window size of 262144 samples,
but did not obtain significantly different results in any case.
The results of this batch Doppler processing are shown in Figure \ref{fig:dopplerscans}.
No visible significant peaks are present,
which indicates that our Doppler processor did not successfully estimate a Doppler peak from the Orion spacecraft.

\section{Summary}

\subsection{Pedagogical outputs}

This project successfully galvanized student excitement for experimental science,
and for space science in particular.
Additionally, students developed practical and transferrable skills, including
\begin{enumerate}
\item Assembly and fielding of a parabolic dish antenna,
\item Azimuth and elevation pointing,
\item Radio system assembly,
\item Basic radio signal processing,
\item Radio data collection, and
\item In-field problem solving.
\end{enumerate}

There were also numerous opportunities for broader education of the public,
including several news reports,
\begin{enumerate}
\item AUNow, American University website: Jack Fredrick, \href{https://www.american.edu/news/fly-me-to-the-moon-for-science.cfm}{Fly Me to the Moon (For Science)}, 2026-04-01
\item WTOP: Jimmy Alexander, \href{https://wtop.com/dc/2026/04/american-university-helps-nasa-track-history}{American University helps NASA track history}, 2026-04-07
\item CAS News, American University website: Patty Housman, \href{https://www.american.edu/cas/news/math-shaping-artemis-ii.cfm}{The Wonder of Artemis II: How Math Is Shaping the Awe-Inspiring Moon Mission}, 2026-04-08
\item WJLA News: Alan Henney, \href{https://wjla.com/news/local/artemis-ii-orion-spacecraft-nasa-artemis-ii-mission-american-university-nasa-project-students-tracking-artemis-spacecraft-orion-spacecraft-radio-signals-nasa-artemis-student-research}{American University students help track NASA’s Artemis II spacecraft}, 2026-04-08
\item NBC4: Beth Brown, \href{https://www.youtube.com/watch?v=Ceo8ZKip4sk}{AU students tracking Artemis II by radio waves holds splashdown watch party}, 2026-04-10
\item WTOP: Tracy Johnke, \href{https://wtop.com/dc/2026/04/american-university-students-celebrate-artemis-iis-safe-return}{`I almost cried': American University students celebrate safe return of Artemis II crew}, 2026-04-10
\item Space Grant: \href{https://spacegrant.org/2026/04/10/dc-space-grant-supports-artemis-ii-student-research-team}{DC Space Grant Supports Artemis II Student Research Team}, 2026-04-10
\end{enumerate}

\subsection{Technical successes}

Even though we did not successfully obtain a Doppler estimate for the Orion spacecraft,
we now have a fully functional end-to-end satellite tracking system.
Additionally, there is a team of students who are trained in tracking operations.
These skills can also be applied to collect radio astronomy data.

\subsection{Future projects}

The adventure has inspired excitement for science and engineering,
and numerous other project ideas are being discussed for the next experiment!
These future goals include downlinking data from CubeSats,
collecting weather maps from geostationary satellites,
measuring the cosmic microwave background,
and hunting for pulsars.

\section*{Acknowledgements}

We would like to thank the students who were involved in helping with this project: Naomi Morris, Shafaq Yousaf, Toby Wieland, Martina Wyland, Isabella Jones.

We would also like to thank the Physics department staff: Phil Johnson and Chelsey Brown.
Joint funding from the American University Physics Department and the NASA District of Columbia Space Grant Consortium is gratefully acknowledged.
We thank Larry Clatterbuck at Airlie Farm for coordinating the installation of our equipment.
We thank Daniel Yochelson for feedpoint bracket manufacturing advice.
Finally, we are grateful to NASA Space Communication and Navigation for giving us the opportunity to participate in this project.

\newpage

\bibliographystyle{plain}
\bibliography{2026_artemis2_bib}

\end{document}